\documentclass[a4paper]{jpconf}
\usepackage{graphicx}
\begin{document}
\title{Cosmic ray acceleration in accretion flows of galaxy clusters}

\author{V.N.Zirakashvili, V.S.Ptuskin}

\address{Pushkov Institute of Terrestrial Magnetism,
Ionosphere and Radiowave Propagation, 108840 Moscow Troitsk, Russia}

\ead{zirak@izmiran.ru}

\begin{abstract}
We investigate acceleration of cosmic rays by shocks and accretion flows in galaxy clusters. Numerical results for
spectra of accelerated particles and nonthermal emission are presented. It is shown that the acceleration
of protons and nuclei
in the nearby galaxy cluster Virgo can explain the observed spectra of ultra high energy cosmic rays.
\end{abstract}



\section{Introduction}

Clusters of galaxies are considered as a possible candidate for
the origin of ultra-high energy cosmic rays (UHECRs) (see e.g.
\cite{brunetti14} for a review). Primordial density fluctuations are
amplified via gravitational instability in the expanding Universe
and result in the appearance of different coherent density structures at
the present epoch. The galaxy clusters are formed at the latest
times and continue to grow presently due to accretion of the 
cicumcluster gas and dark matter.  The inflow of matter is
accompanied by an accretion shock at a distance several Mpc from the
cluster center.

Virgo cluster of galaxies with the total mass of $\sim 10^{15}M_{\odot }$ at the distance  $d\sim 15-20$ Mpc \cite{ade16}
from the Milky Way is the nearest large galaxy cluster.
It is located at the center of the Local super cluster of galaxies. Because of its proximity Virgo has been proposed 
 as the source
of observed UHECRs within a phenomenological diffusive model \cite{giler80}.

The diffusive shock acceleration (DSA) process \cite{krymsky77,bell78,axford77,blandford78} is believed to be 
 the principal mechanism for the production of
galactic cosmic rays (CR) in supernova remnants (SNRs). Over the last decade
the excellent results of X-ray and gamma-ray astronomy have been providing evidence 
of the presence of multi-TeV energetic
particles in these objects (see e.g. \cite{lemoine14}).

Since the accretion shocks are very large they can accelerate particles to significantly higher energies
 compared to SNRs and produce a significant fraction of observable UHECRs \cite{norman95,kang96,kang97}.
In this paper we describe the modifications of our non-linear
 DSA model \cite{zirakashvili12} designed for the investigation of DSA in SNRs and apply
 it for the investigation of particle acceleration in clusters of galaxies. In particular we apply the model for the
 Virgo cluster of galaxies.
%

\section{Nonlinear diffusive shock acceleration model}

Details of our model of nonlinear DSA can
be found in \cite{zirakashvili12}. The model contains coupled
spherically symmetric hydrodynamic equations  and the transport
equations for energetic protons, ions and electrons.

The hydrodynamical equations for the gas density  $\rho (r,t)$, gas velocity $u(r,t)$,
gas pressure
$P_g(r,t)$, magnetic pressure $P_m(r,t)$, and the equation for isotropic part of the cosmic ray proton momentum
distribution
 $N(r,t,p)$, the momentum per nucleon  distribution $N_i(A, r,t,p)$ of nuclei with atomic number $A$
and dark matter velocity distribution $F(r,t,w)$
in the spherically symmetrical  case are given by

\begin{equation}
\frac {\partial \rho }{\partial t}=-\frac {1}{r^2}\frac {\partial }{\partial r}r^2u\rho
\end{equation}

\begin{equation}
\frac {\partial u}{\partial t}= g(r)-u\frac {\partial u}{\partial r}-\frac {1}{\rho }
\left( \frac {\partial P_g}{\partial r}+\frac {\partial P_c}{\partial r}
+\frac {\partial P_m}{\partial r}\right)
\end{equation}

\begin{eqnarray}
\frac {\partial P_g}{\partial t}+u\frac {\partial P_g}{\partial r}
+\frac {\gamma _gP_g}{r^2}\frac {\partial r^2u}{\partial r}= -(\gamma _g-1)\Lambda _c(T_e)n^2
\end{eqnarray}

\begin{eqnarray}
\frac {\partial P_m}{\partial t}+u\frac {\partial P_m}{\partial r}
+\frac {\gamma _mP_m}{r^2}\frac {\partial r^2u}{\partial r}=0
\end{eqnarray}

\begin{eqnarray}
\frac {\partial N}{\partial t}=\frac {1}{r^2}\frac {\partial }{\partial r}r^2D(p,r,t)
\frac {\partial N}{\partial r}
-u\frac {\partial N}{\partial r}+\frac {\partial N}{\partial p}
\frac {p}{3r^2}\frac {\partial r^2u}{\partial r}
+\frac 1{p^2}\frac {\partial }{\partial p}p^2b(p)N+4\nu _{ph}(4)N_i(4)+\sum _{A=5}^{56}\nu _{ph}(A)N_i(A)
\end{eqnarray}
\begin{eqnarray}
\frac {\partial N_i(A)}{\partial t}=\frac {1}{r^2}\frac {\partial }{\partial r}r^2D_i(p,r,t)
\frac {\partial N_i(A)}{\partial r}
-u\frac {\partial N_i(A)}{\partial r}
+\frac {\partial N_i(A)}{\partial p}
\frac {p}{3r^2}\frac {\partial r^2u}{\partial r}
+\frac 1{p^2}\frac {\partial }{\partial p}p^2b(p)N_i(A)\nonumber \\
-\nu _{ph}(A)N_i(A)+\nu _{ph}(A+1)N_i(A+1)
\end{eqnarray}
\begin{eqnarray}
\frac {\partial F}{\partial t}=-\frac {1}{r^2}\frac {\partial }{\partial r}r^2wF
-g(r)\frac {\partial F}{\partial w}
\end{eqnarray}
Here $P_c=4\pi \int dpp^3v(N+\sum _AAN_i(A))/3$ is the cosmic ray pressure, 
$T_e$, $\gamma
_g$ and $n$ are the gas temperature, adiabatic index and number
density respectively,  $\gamma _m$ is the magnetic adiabatic index, $D(r,t,p)$ is the cosmic ray diffusion
coefficient. The radiative cooling of gas is described by the
cooling function $\Lambda _c(T_e)$.

The gravitational acceleration $g(r)$ is given by the following expression
\begin{equation}
g(r)=\Lambda r-\frac {4\pi G}{r^2}\int _0^{r}r_1^2dr_1(\rho (r_1)+\rho _{DM}(r_1))
\end{equation}
where $G$ is the gravitational constant, $\Lambda $ is the lambda term and $\rho _{DM}(r)=\int dw F(r,w)$ is
the dark matter density. 

The function $b(p)$ describes the energy losses of particles. In
particular the losses of multi EeV protons and
nuclei via electron-positron pair production on the
microwave background radiation (MWBR) 
  are important in galaxy clusters. Another losses of energy of the nuclei occur in  photonuclear reactions with
infrared photons. They are described by two last terms in Eq. (6). We use a simplified approach and
 assume that the nuclei lose one nucleon in every photonuclear reaction. The corresponding proton source term
is added in Eq. (5). For the calculation of photonuclear cross-sections we use the background spectrum
of infrared photons
$n(\epsilon )=0.87\cdot 10^{-3}a^{-4}(t)\epsilon ^{-2}$ cm$^{-3}$eV$^{-1}$, where $\epsilon $ is the photon energy expressed 
 in eV and $a(t)$ is the expansion factor 
 of the Universe. At the present epoch ($a=1$) the spectrum  
is three times lower than LIR spectrum
 of Puget et al. \cite{puget75} and is close to the lower limit of Malkan \& Stecker \cite{malkan98}. 

Cosmic ray diffusion is determined by particle scattering on
magnetic inhomogeneities. Here we consider the simplest model with
strong background random magnetic fields in the circumcluster medium.
The properties of the gas and magnetic field in the circumcluster
medium are determined by galactic winds of individual galaxies.
These winds were strong in past when the star formation rate was
significantly higher than at the present epoch. The winds
 were driven by cosmic rays and hot gas produced by supernova explosions. The gas was heated,
galactic cosmic rays
were reaccelerated and magnetic fields were amplified at termination shocks of
the galactic winds. That is why we expect that these three components have comparable
energy densities in the circumcluster medium. Additionally the galactic winds transport
 metals and therefore the metallicity of the circumcluster medium is different from the
 primordial one that is indeed  observed.

Bearing in mind that the magnetic field is strong in the circumcluster
medium we neglect its generation via cosmic ray streaming
instability in Eq. (4). We also neglect the gas heating via magnetic
dissipation in  Eq. (3). So the gas and magnetic pressures $P_g$ and
$P_m$ evolve in accordance with adiabatic compression of the gas.
Below we use the magnetic adiabatic index $\gamma _m=3/2$.

We use Bohm-like dependence of the diffusion coefficient $D$
 of energetic particles with the charge $q$, momentum $p$ and speed $v$

\begin{equation}
D=\eta _BD_B, \ D_B=\frac {cpv}{3qB}, \ B=\sqrt{8\pi P_m}
\end{equation}
where $B$ is the total magnetic field strength, while $\eta_B$ is
 a free parameter characterizing speed of diffusion in the terms of Bohm diffusion. 
Since the Bohm diffusion coefficient $D_B$ is the
minimal possible value we shall use the value of $\eta _B=2$ below.

 In the shock transition region the magnetic pressure
 is increased by a  factor of $\sigma ^{\gamma _m}$,  where
$\sigma $ is the shock compression ratio. Its impact  on the shock
dynamics is taken into account via the Hugoniot conditions.

We do not consider the injection of thermal ions at the shock front in the present study.
As was mentioned before there are background cosmic rays in the circumcluster medium. We assumed that
 the spectrum of these particles is given by

\begin{equation}
N_{IG}(p)=N_0 \left( \frac {a(t)p}{p_{IG}}\right) ^{-\gamma -2}\exp{\left( -\frac {a(t)p}{p_{IG}}\right) }
\end{equation}
Here $p_{IG}$ is the maximum momentum of background cosmic rays at the present epoch and 
 $\gamma $ is the spectral index of differential energy distribution of background cosmic rays.

The similar expression was used for the background spectra of
nuclei. Helium nuclei were taken 10 times less abundant than
protons. For heavier nuclei we assumed an enrichment proportional to the atomic number 
$A$ as observed in Galactic cosmic rays.

We neglect the pressure of energetic electrons and treat them as test particles.
The evolution of the
electron distribution is described by the equation analogous to Eq. (5)
with the function $b(p)$  describing Coulomb, synchrotron and inverse Compton
(IC) losses and additional terms describing the production of secondary
leptons by energetic protons and nuclei.

The following expressions were used for the  circumcluster magnetic and gas pressures
\begin{equation}
P_{gIG}=a(t)^{-3\gamma _g}P_{g0}, \ P_{mIG}=a(t)^{-3\gamma _m}P_{m0},
\end{equation}
where $P_{g0}$ and $P_{m0}$ are the gas and magnetic circumcluster pressures at the present epoch respectively.

\section{Modeling of DSA in the Virgo cluster}

We model the Virgo cluster formation and particle acceleration for the cosmological parameters
$\Omega _b=0.05$, $\Omega _{DM}=0.25$,
$\Omega _\Lambda=0.7$ and the Hubble constant $H=70$ km s$^{-1}$ Mpc$^{-1}$. The corresponding mean barion density 
 $\rho _b$ of the  Universe is $\rho _b=3\Omega _bH^2/8\pi G=4.6\cdot 10^{-31}$g cm$^{-3}$. The lambda term in  Eq. (8) 
is given by $\Lambda =\Omega _{\Lambda }H^{2}$. The initial instant of time is $t_0=4$ Gyr from the Big Bang.

The parameters of the background cosmic ray spectrum given by Eq. (10) were adjusted to reproduce
observations of Auger Collaboration. We used the values  $\gamma =1.8$ and $p_{IG}=8$ PeV/c. The spectra of background
 cosmic rays are somewhat harder than the expected spectra of galactic sources. However the additional
reacceleration at galactic wind termination shocks indeed will make galactic cosmic ray spectra harder.

We use the initial condition corresponding to the self-similar solution of Bertschinger \cite{bertschinger85}. 
At latest times the accelerated expansion of the Universe results in the higher accretion shock speed and radius 
in comparison with this solution. 

\begin{figure}[h]
\includegraphics[width=9.0cm]{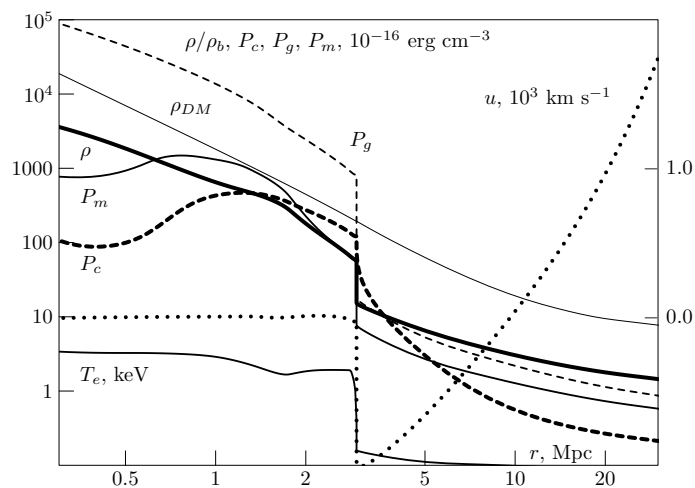}\hspace{2pc}%
\begin{minipage}[b]{14pc}\caption{The radial dependence of the gas density (thick solid line), dark matter density 
(thin dotted line), the gas
velocity (dotted line), cosmic ray pressure (thick dashed line),
the magnetic pressure $B^2/8\pi =P_m$ (solid line),
the gas
temperature $T_e$ (thin solid line)
and the gas pressure $P_g$ (dashed line) at
 $t=13.5$ Gyr in the Virgo cluster. }
\end{minipage}
\end{figure}

Figures (1)-(5) illustrate the results of our numerical calculations.

\begin{figure}[h]
\includegraphics[width=9.0cm]{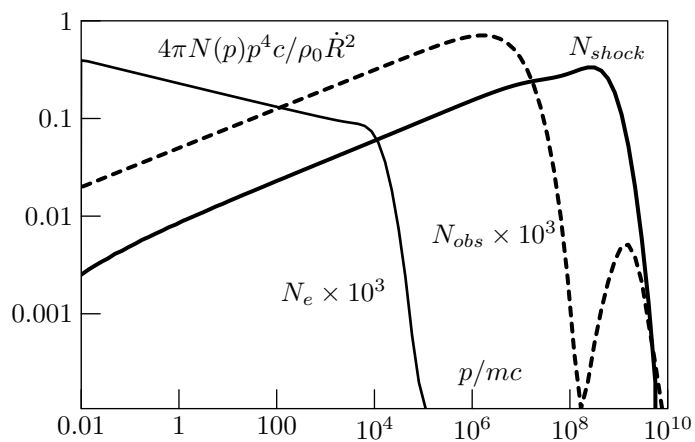}\hspace{2pc}%
\begin{minipage}[b]{14pc}\caption{Spectra of accelerated particles at the accretion shock of Virgo cluster.  The
spectrum of protons (thick solid line) and electrons (thin solid line) are shown.
We  also show the proton spectrum at the Milky Way position at $r=20$ Mpc (dashed line). }
\end{minipage}
\end{figure}

The radial dependence of physical quantities in the Virgo cluster at the present epoch 
 ($t=13.5$ Gyr) are shown in Fig.1. At large distances we have the uniform expansion of the Universe.
 Closer to the cluster deviations from the Hubble law due to the initial local overdensity occur. 
We observe an inflow of
 matter inside a so-called "turnaround' point at $r\approx $10 Mpc.
 The accretion shock is situated at $r=R=3$ Mpc. The gas temperature is several keV inside the shock. The calculated 
gas density 
 profile is in agreement with observations of the Virgo cluster \cite{ade16}. The total mass of the gas inside the shock  
 1.5$\cdot 10^{14}M_{\odot }$ is also close to the observed value \cite{ade16}.

The cosmic ray, magnetic and gas  pressures at large distances are of the order of $10^{-16}$ erg cm$^{-3}$ that is
 considered as a characteristic value for the extragalactic pressure around the Milky Way. The corresponding strength of
 the background magnetic field is $B_{IG}=3\cdot 10^{-8}$G that is a reasonable value for the Local supercluster.

It is interesting that the adjusted level of background cosmic rays results in the pressure of reaccelerated particles
 close to $10\% $ of the ram pressure of the accretion shock. This number is similar to the one found in the hybrid
modeling of collisionless quasiparallel shocks \cite{caprioli14}  where the ions are injected at the shock front.

\begin{figure}[h]
\includegraphics[width=9.0cm]{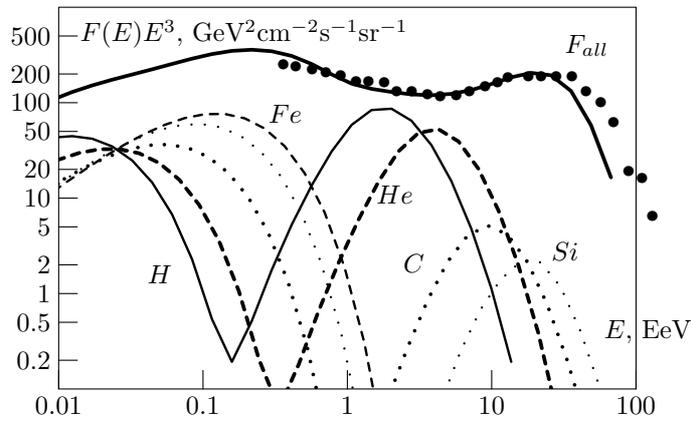}\hspace{2pc}%
\begin{minipage}[b]{14pc}\caption{Calculated all-particle spectrum  (thick solid curve),
the spectrum  of protons (thin solid line) and the spectra
of other nuclei at the Milky Way position.
Observational data of Auger Collaboration \cite{auger13} (circles) are also shown. }
\end{minipage}
\end{figure}

\begin{figure}[h]
\includegraphics[width=9.0cm]{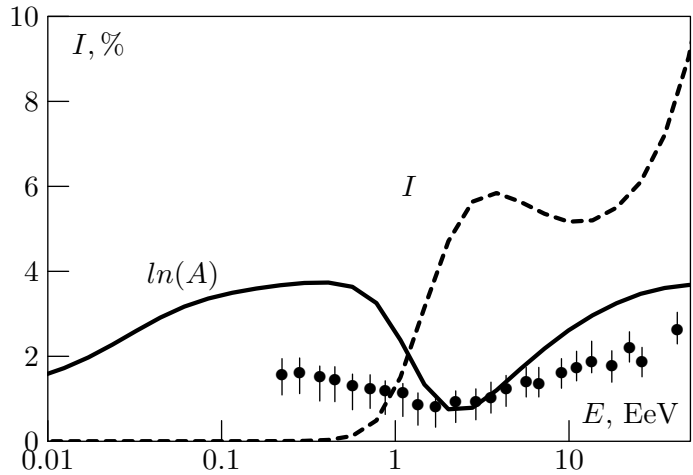}\hspace{2pc}%
\begin{minipage}[b]{14pc}\caption{Calculated mean logarithm (solid line) and cosmic ray anisotropy (dashed line) 
at the Milky Way position.
 Experimentally measured mean logarithm
data of Auger collaboration \cite{auger13} using EPOS-LHC model of
particle interactions in the atmosphere (data with error-bars) are
also shown.  }
\end{minipage}
\end{figure}

Spectra of accelerated in Virgo protons and electrons at $t=13.5$ Gyr
are shown in Fig.2. At this point the maximum energy of accelerated
protons is close to $10^{18}$ eV. Only highest energy protons and nuclei from the cut-off region reach
 the Milky Way. Lower energy particles simply have not time to propagate from
 the shock surface to the Milky Way. We can observe only exponentially small
fraction of protons ($10^{-4}$) reaccelerated at the accretion shock. So the particles are in a rather
unusual propagation
 mode (see also \cite{kang96}). At lower energies we observe the  background circumcluster cosmic rays.

\begin{figure}[h]
\includegraphics[width=9.0cm]{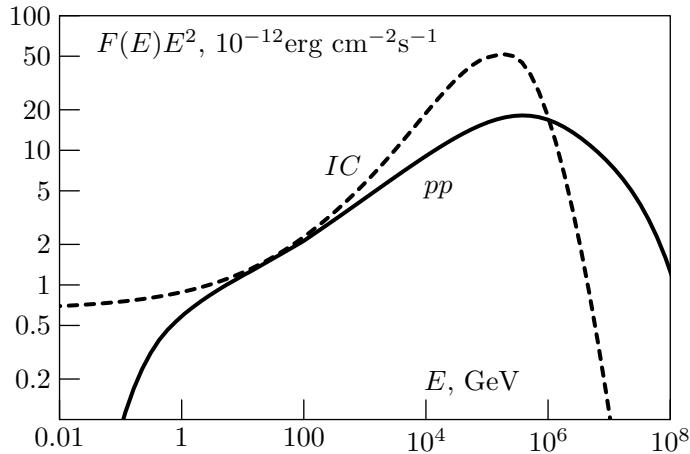}\hspace{2pc}%
\begin{minipage}[b]{14pc}\caption{The results of modeling of gamma radiation of Virgo.
We show IC emission
(dashed line) and  gamma-ray emission from pion decay (solid line). The absorption  of  multi-TeV photons 
due to pair production is not taken into account. }
\end{minipage}
\end{figure}

All-particle spectrum and spectra of individual nuclei calculated
for the Milky Way position ($r=20$ Mpc) are shown in Fig.3. The ankle at 5 EeV
corresponds to the transition from the proton-Helium composition to
the heavier nuclei dominated composition in our model. The heaviest
nuclei do not present in the spectrum because of the photonuclear
disintegration. The cut-off energy of the spectrum is mainly
determined by the photonuclear reactions of heavy nuclei during
acceleration and propagation. If we neglect the photonuclear
reactions the maximum energy will be several times higher.

Calculated mean logarithm of atomic number and anisotropy are shown in Fig.4. The model reproduces light composition
 at EeV energies and the dipole anisotropy below the value of $6.5$ percent measured by Auger Collaboration \cite{auger17}.

The spectra of unabsorbed  gamma rays expected from the Virgo cluster  are shown
in Fig. 5. The gamma emission shows two components. One is
produced by the nuclear collisions of protons and nuclei  with gas
in the cluster interior while another one is the
 inverse Compton component that at TeV energies 
is produced by electron positron pairs generated via interaction of
accelerated particles with background photons of MWBR. The last process is rather efficient in galaxy clusters
\cite{vannoni09, inoue08}. In spite of the high expected flux the large angular size of the source (several degrees) makes the gamma 
 detection of the Virgo cluster very difficult. 

\section{Discussion}

We show that almost all UHECRs can be accelerated in the nearby Virgo cluster of galaxies. Our consideration is close to the
 earlier pure proton model of Kang et al. \cite{kang96}. Note that we used more realistic value of 
the magnetic field strength that is 10 times lower. The corresponding decrease of the maximum energy is 
compensated by the presence of heavier nuclei in UHECR spectrum. 

The drawback of the model is its sensitivity to adjusted parameters to fit UHECR data. As mentioned before we observe 
 an exponentially small part (10$^{-4}$) of the highest energy particles accelerated at the accretion shock. 
That is why modest variations of the magnetic field strength result in the significant change of UHECR intensity. 
 
Probably more massive
 and more distant clusters like Coma or another astrophysical sources (e.g. active galactic nuclei)
also might give some contribution at highest energies.
The particles accelerated in distant astrophysical objects can reach the Local supercluster propagating
in the cosmological voids where the magnetic fields are very weak.

The adjusted maximum energy of background cosmic rays $E_{IG}=cp_{IG}= 8 \ Z$ Pev is not far from the "knee" energy.
It is not excluded that this is not a coincidence.
The value of $E_{IG}$ is rather well constrained by available UHECR data. For lower values of $E_{IG}$ we would observe
 a large dip in the spectrum between 0.1 and 1 EeV. For higher values of $E_{IG}$ we would observe heavy UHECR
 composition at 1 EeV.

\ack The work was supported by Russian Foundation of Fundamental
Research grant 16-02-00255. The paper was presented at the 26th
European Cosmic Ray Symposium, Barnaul, July 6 - 10, 2018.

\bigskip 

\begin{thebibliography}{99}   


\bibitem{brunetti14} Brunetti, G., \& Jones, T., 2014, Intern. Journal of Modern Physics D, 29, 3007

\bibitem{ade16} Ade, P.A.R., Aghanim, N., Arnaud, M. et al., 2016, A\&A, 596, 101


\bibitem{giler80} Giler, M., Wdowczyk, W., \& Wolfendale, A.W., 1980, J. Phys. G, 6, 1561

\bibitem{krymsky77} Krymsky, G.F. 1977, Soviet Physics-Doklady, 22, 327

\bibitem{bell78} Bell, A.R., 1978, MNRAS, 182, 147

\bibitem{axford77} Axford, W.I., Leer, E. \& Skadron, G., 1977, Proc. 15th
ICRC, Plovdiv, 90, 937

\bibitem{blandford78} Blandford, R.D., \& Ostriker, J.P. 1978, ApJ, 221, L29

\bibitem{lemoine14} Lemoine-Goumard M. 2004, Proc.  IAU Symp., 296, 287

\bibitem{norman95} Norman, C.A., Melrose, D.B., Achterberg, A., 1995, ApJ, 454, 60

\bibitem{kang96} Kang, H., Ryu, D., Jones, T.W., 1996, ApJ, 456, 422

\bibitem{kang97} Kang, H., Rachen, J.P., Biermann, P.L., 1997, MNRAS, 286, 257


\bibitem{zirakashvili12} Zirakashvili, V.N. \& Ptuskin V.S. 2012, Astropart. Phys., 39, 12




\bibitem{puget75}  Puget, J.L.,  Stecker, F.W., \&  Bredekamp, J.H., 1975, ApJ, 205, 638

\bibitem{malkan98} Malkan, M.A., \& Stecker, F.W., 1998, ApJ, 496, 13

\bibitem{bertschinger85} Bertschinger, M., 1985, ApJS, 58, 39 

\bibitem{caprioli14} Caprioli, D., 2014, Nuclear Physics B (Proc. Suppl.), 256, 48

\bibitem{auger13} Letesser-Selvon, A. et al.., Highlights from the Pierre Auger Observatory,
Proc. 33th ICRC, Rio de Janeiro 2013,  arXive:1310.4620

\bibitem{auger17} Pierre Auger Collaboration, 2017  arXive:1709.07321

\bibitem{vannoni09} Vannoni, G., Aharonian, F.A., Gabici, S., Kelner, S.R., Prosekin, A., 2011,
A\&A, 536,56

\bibitem{inoue08} Inoue, S., Sigle, G., Miniati, F., \& Armengaud, E., 2008, Proc. 30th ICRC, 4, 555



\end{thebibliography}

\end{document}